# High-resolution and gapless dual comb spectroscopy with current-tuned quantum cascade lasers

MICHELE GIANELLA,[1,*] AKSHAY NATARAJ,[1] BÉLA TUZSON,[1] PIERRE JOUY,[2] FILIPPOS KAPSALIDIS,[3] MATTIAS BECK,[3] MARKUS MANGOLD,[2] ANDREAS HUGI,[2] JÉRÔME FAIST,[3] AND LUKAS EMMENEGGER[1]

[1]*Empa, Laboratory for Air Pollution & Environmental Technology, 8600 Dübendorf, Switzerland*
[2]*IRsweep AG, Laubisrütistrasse 44, 8712 Stäfa, Switzerland*
[3]*Institute for Quantum Electronics, ETH Zurich, 8093 Zurich, Switzerland*
*\*Corresponding author: michele.gianella@empa.ch*

**Abstract:** We present gapless, high-resolution absorption and dispersion spectra obtained with quantum cascade laser frequency combs covering 55 cm$^{-1}$. Using phase-sensitive dual comb design, the comb lines are gradually swept over 10 GHz, corresponding to the free spectral range of the laser devices, by applying a current modulation. We show that with interleaving the spectral point spacing is reduced by more than four orders of magnitude over the full spectral span of the frequency comb. The potential of this technique for high-precision gas sensing is illustrated by measuring the low pressure (107 hPa) absorption *and* dispersion spectra of methane spanning the range of 1170 cm$^{-1}$ - 1225 cm$^{-1}$ with a resolution of 0.001 cm$^{-1}$.

## 1. Introduction

Recent developments in optical frequency combs (FCs), especially in the mid-infrared (MIR) spectral region, promise improved molecular sensing for industrial, environmental and biomedical applications [1,2]. Frequency combs with emission in the MIR spectral region – where many relevant molecules possess strong and characteristic fundamental ro-vibrational absorption bands [3–6] – include sources based on non-linear mixing with near-infrared femtosecond fiber lasers [7–11], microresonator-based approaches [12,13], interband cascade lasers [14], and quantum cascade lasers (QCLs) [15,16]. The latter two are of special interest because they are compact, robust, and they provide high optical power. Due to the small footprint and modest power requirement, these semiconductor devices have significant potential for the miniaturization of frequency comb spectrometers. This is somewhat in contrast to indirect MIR frequency comb sources relying e.g. on difference frequency generation with fiber-laser combs, which tend to be significantly more complex and bulky [17–20].

In terms of spectroscopy, the absorption coefficient and refractive index of the sample are encoded in the magnitude and phase of the comb lines of the interrogating comb [21–27]. While there are several techniques for magnitude and phase retrieval [2], the present work relies on dual comb spectroscopy (DCS), which is emerging as a powerful technique for fast and sensitive broadband molecular spectroscopy. It allows for fast acquisition over the full spectral coverage of the comb and high resolving power without any moving parts [28]. In DCS, the *interrogating* comb is used to probe the sample, while the *local oscillator* comb, with a slightly different comb spacing, is used to produce an *interferogram* which arises from the multi-heterodyning of the various comb lines. In the *phase-sensitive* configuration, the beam of the local oscillator bypasses the sample, but often the sample is probed with both interrogating and local oscillator beams and the sample's dispersion cannot be measured.

The discrete Fourier transform of the interferogram (the *multi-heterodyne signal*) consists of a number of evenly spaced narrow-band components called *beat notes*, each of which carries the information about the magnitude product (or power) and phase difference of the two optical comb lines that generated it. A general pre-requisite for DCS to work is that the observation time during which the interferogram is acquired must be long enough to permit the individual beat notes to be resolved and, at the same time, short enough to maintain the beat note linewidth below the spacing of adjacent beat notes. While most examples of DCS employ mutually-locked laser sources, it is perfectly possible to operate a spectrometer with two free-running QCL-FCs [21]. The reason this works well lies in the small intrinsic linewidth and large free spectral range (FSR) – or repetition rate – of quantum cascade lasers [29], which results in sufficient mutual coherence of the two lasers even if they are unlocked. With 200 comb lines and a detection bandwidth of 1 GHz, the beat notes can be spaced up to 5 MHz. Since mutual coherence times of a few tens of μs can be readily achieved, the aforementioned conditions are thus fulfilled.

For every spectrometer, there is a *spectral point spacing* (spectral distance between adjacent points in a spectrum) and a *spectral resolution* (frequency difference of two lines in a spectrum that are still distinguishable). Regardless of the frequency comb type, there is a large mismatch between the point spacing and the resolution, with the point spacing being several orders of magnitude larger than the resolution. This is true for QCL-FCs as well, where point spacing is typically $5 - 15$ GHz, given by the FSR of the device, and the resolution is of the order of 1 MHz. Whilst ~10 GHz point spacing may be sufficient for spectroscopy in the condensed phase [27], gas-phase measurements of narrow absorption features typically require a much denser spectral point spacing. By interleaving of multiple offset spectra the point spacing can be reduced significantly [30,31].

Here, we present gapless spectra (i.e., with point spacing smaller than the resolution) obtained with free-running QCL-FCs using dual comb spectroscopy. Through interleaving, the spectral point spacing is reduced by more than four orders of magnitude from 9.8 GHz (FSR of the lasers) down to 300 kHz over the full 55 cm$^{-1}$ (1.7 THz) span of the interrogating comb, thus paving the way for multi-species gas detection with QCL-FCs with sub-second time resolution.

## 2. Experimental realization

The dual-comb spectrometer is based on the phase-sensitive (dispersive) design with current modulation applied on both lasers, as shown in **Fig. 1**(a). The setup uses an IRis-F1 spectrometer (IRsweep AG, Switzerland) that has been modified with respect to laser driving, optics, and data processing. The two laser sources are InGaAs/AlInAs on InP-based dual-stack QCLs [32]. They are 4.5 mm long (FSR ~ 9.8 GHz, FSR difference ~ 4.5 MHz) and have both facets uncoated, giving up to 400 mW of optical power at room temperature per facet. The thermo-electrically cooled HgCdTe sample photodetector (IRsweep AG) measures the interferogram of the two combs, whereby the interrogating comb passes through the sample; the normalizing photodetector serves to cancel or suppress common mode fluctuations due to power and frequency instabilities. The detector signals are bandpass filtered (ZX75-LP-1050-S+, MiniCircuits) and then digitized at a sampling rate of 2 GSa/s with 14 bit resolution. After each sweep, the raw data are transferred to a hard disk drive without any initial processing.

The comb line frequencies are swept by applying a triangular current modulation to the lasers (see **Fig. 1**(b)). To prevent the beat note frequencies from moving outside of the detection bandwidth (1 GHz), the drive currents of *both* lasers are modulated simultaneously. As the QCL-FCs have slightly different tuning properties, the two triangular waveforms have different slopes and offsets, and are chosen such that they minimize the offset and breathing of the multi-heterodyne signal during the sweep. The acquisition takes place during the rising segment of the triangle, which lasts 120 ms. The falling segment lasts 800 ms, and the time delay between successive sweeps is thus 920 ms. Both waveforms are set to start on a trigger pulse which also starts the acquisition. On each trigger pulse, $2^{28}$ data points per channel are acquired at 2 GSa/s and saved to disk to be processed later. Two series of sweeps are measured. In the first, the sample cell is removed from the laser beam path (*background* measurement), and in the second the cell is placed into the path (*sample* measurement).

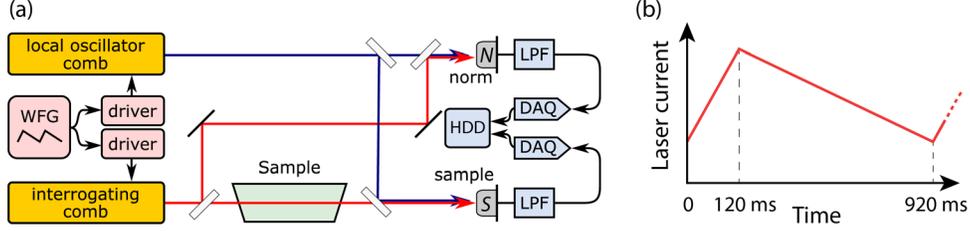

**Fig. 1**. (a) Scheme of the dual comb spectrometer. WFG: waveform generator, sample/norm: photodetectors, LPF: low-pass filter, DAQ: data acquisition board, HDD: hard disk. (b) Laser current modulation waveform.

After the acquisitions have been completed, the raw data of each sweep are divided into slices of 16.4 µs duration, apodized with a flat-top window function, and fast-Fourier-transformed (FFT). The power spectrum of a typical slice (i.e., the magnitude squared of the Fourier coefficients) is shown in **Fig. 2**. The peaks correspond to the beat notes generated by the heterodyning of the various comb lines of the two lasers. Each beat note is generated by a line from the interrogating comb (with frequency $\nu_{js}$) and a line from the local oscillator comb ($j$: beat note index, $s$: slice (time) index). We define the *complex beat note amplitudes*, $S_{js}$, as the values of the Fourier coefficients at each local maximum in **Fig. 2**. The so-defined beat note *amplitudes* are sensitive to fluctuations in the beat note *frequencies* – an undesired property – because frequency fluctuations tend to smear the beat note power across multiple Fourier coefficients. In addition, *spectral leakage* [33], a consequence of the finite measurement duration (in our case, 16.4 µs), introduces a frequency-dependent distortion of both the magnitude and phase of the amplitudes. The consequence of these two effects is that the *measured* beat note amplitude fluctuates very strongly if the frequency of the beat note varies, even if the *true* beat note amplitude is perfectly constant. The purpose of the normalization detector in **Fig. 1**(a) is to provide a reference measurement, $N_{js}$, for all the beat note amplitudes, $S_{js}$, measured on the sample detector so that a complex amplitude *ratio*,

$$R_{js} = S_{js}/N_{js}, \quad (1)$$

can be computed (all beat notes that appear on the sample detector are also present on the normalizing detector). The ratio was found to exhibit much smaller fluctuations than the beat note amplitudes. This is explained by the fact that both spectral leakage and the smearing effect described above affect both measurements in the same way, leading to a strong correlation between them.

In order to compute the absorbance and phase-shift of a sample, the amplitude ratio of each beat note needs to be measured two times. One measurement with the sample in the beam path; and another measurement when the sample cell is removed (or emptied). This second *background* measurement serves to normalize the emission spectrum of the sources and cancel the instrumental attenuation and phase-shift introduced by the various optical elements and propagation distances. If $R_{js}^{\mathrm{smp}} = S_{js}^{\mathrm{smp}}/N_{js}^{\mathrm{smp}}$, represents the complex beat note amplitude ratio of beat note $j$ measured on the sample ($S$) and normalizing ($N$) detector, respectively, in time-slice $s$, and $R_{js}^{\mathrm{bkg}} = S_{js}^{\mathrm{bkg}}/N_{js}^{\mathrm{bkg}}$ represents the same amplitude ratio but measured with the sample removed, then the transmittance $T_{js}$ and phase-shift $\psi_{js}$ of the sample are given by the ratio,

$$\sqrt{T_{js}}\exp(i\psi_{js}) = \frac{R_{js}^{\text{smp}}}{R_{js}^{\text{bkg}}}. \tag{2}$$

The absorbance, $A_{js}$, is given by

$$A_{js} = -\ln(T_{js}) = 2\ln\left(\left|R_{js}^{\text{bkg}}\right|\right) - 2\ln\left(\left|R_{js}^{\text{smp}}\right|\right), \tag{3}$$

and the phase-shift, $\psi_{js}$, by

$$\psi_{js} = \arg(R_{js}^{\text{smp}}) - \arg\left(R_{js}^{\text{bkg}}\right). \tag{4}$$

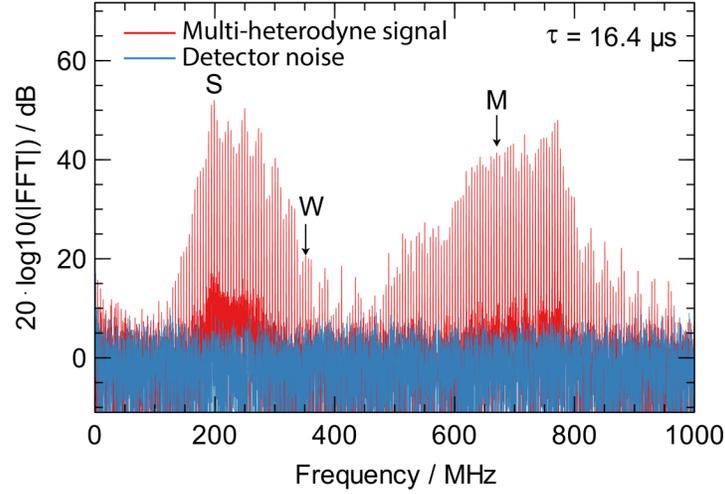

**Fig. 2.** Spectrum of a 16.4 µs long interferogram. The root-mean-square of the detector noise floor is set to 0 dB. Strong (S), medium (M), and weak (W) beat notes used later in *Noise characterization* are indicated.

## 3. Results and discussion

*Fully interleaved methane absorption and dispersion spectra*

For demonstrating the capability of our spectral interleaving approach, low pressure (107 hPa) gas phase absorption and dispersion spectra of methane are measured. At this pressure, the full width at half maximum of the methane absorption lines is approximately 400 MHz. In **Fig. 3**, the measured spectra are shown along the absorption spectrum simulated using HITRAN data [34]. For the *sample* measurement, 22 sweeps (22 × 120 ms) were co-averaged for the darker trace and one sweep (120 ms) for the lighter trace, whereas for the *background* 50 sweeps were co-averaged in both cases with the cell removed from the path. Each spectrum contained over 5 million data points with a spectral point spacing of 300 kHz. The data were then decimated by applying a moving average of 100 points, decreasing baseline noise and reducing the spectral resolution to 30 MHz ($0.001\ \text{cm}^{-1}$). The sinusoid in **Fig. 3** is an interference fringe with a period of $0.62\ \text{cm}^{-1}$ adequately shows that the absorbance traces of the various comb lines, covering approximately $0.33\ \text{cm}^{-1}$ each, have been stitched together correctly. As there is little optical power available in the spectral region between $1190\ \text{cm}^{-1}$ and $1200\ \text{cm}^{-1}$, this led to significantly more noise in that region than in the rest of the spectrum, but co-averaging of multiple sweeps strongly reduces the noise in this region as well. In the dispersion spectrum, a linear trend, attributed to the dispersion of the gas cell windows, was removed from the phase data.

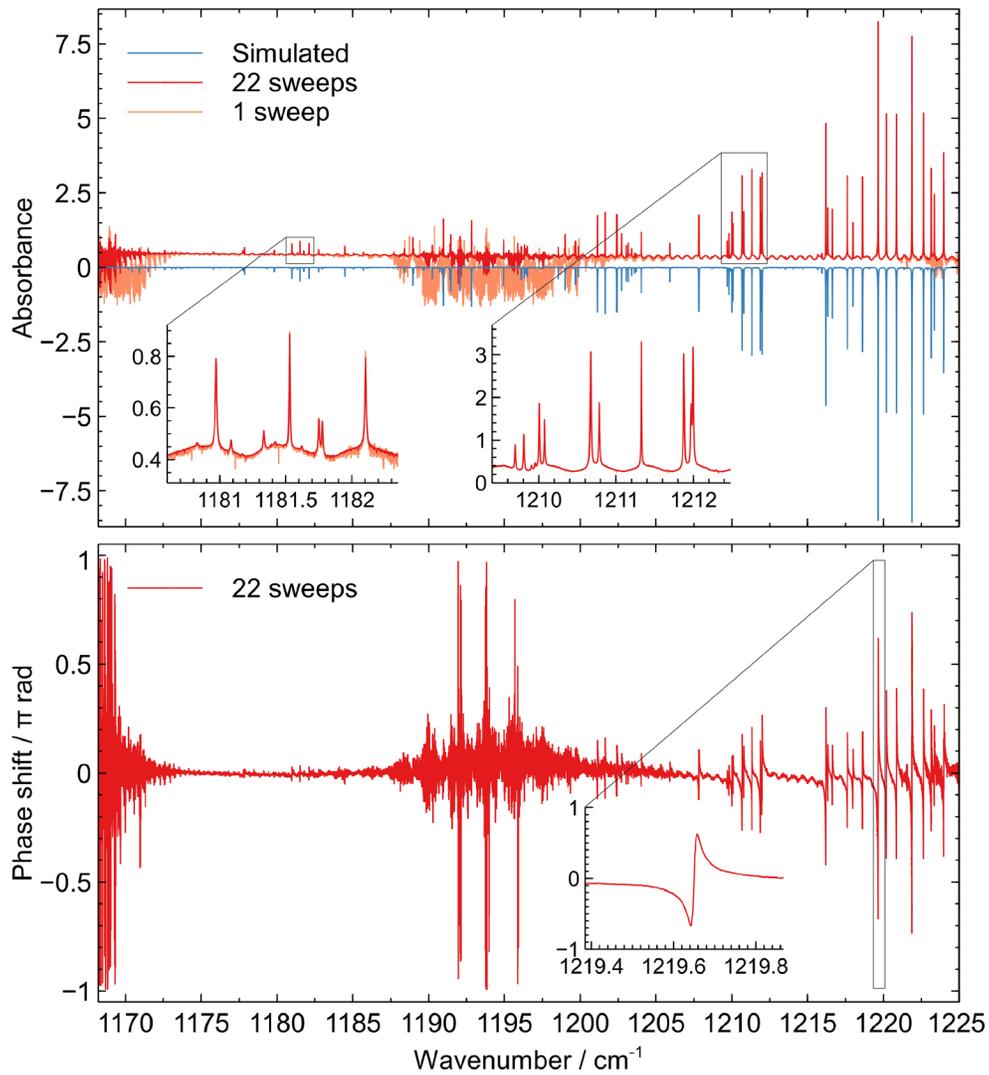

**Fig. 3.** Absorption (top) and dispersion (bottom) spectra of methane. The insets show zooms of the data in the highlighted regions. Conditions: pressure: 107 hPa, path length:14 cm, spectral resolution: 0.001 cm$^{-1}$.

**Fig. 4** shows a Voigt fit to the methane transition near 1209.8 cm$^{-1}$ with a full width at half maximum of ~400 MHz. The root mean square (rms) value of the residuals is $7.4 \times 10^{-3}$ (absorbance units) and, apart from the usual w-shaped residual, there does not appear to be any major systematic deviation from the Voigt line shape. This demonstrates that the retrieval of absorption line shapes with widths of a few hundred MHz is feasible.

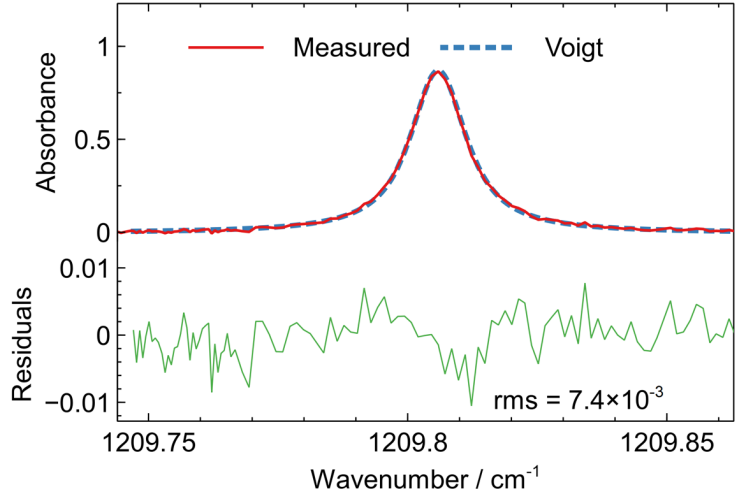

**Fig. 4.** Voigt fit to the CH$_4$ transition at 1209.8 cm$^{-1}$ and associated residuals.

*Frequency axis calibration*

Since the lasers are unreferenced, the frequencies of the comb lines are not known *a priori*. However, it is relatively simple to compute a frequency axis if several absorption lines with known centre frequencies have been measured. Consider the absorbance matrix in Eq. 3, with $j$ as the row index and $s$ as the column index. The entry at row $j$ and column $s$ is the sample absorbance at the optical frequency $v_{js}$. We must now find the function, $f(j, s)$, which, given the comb line index, $j$, and the time slice, $s$, returns the corresponding optical frequency, $v_{js}$. To that end, we exploit the fact that $v_{js}$ are the frequencies of a comb, and thus obey: $v_{js} = v_{j_0 s} + (j - j_0) \cdot \Delta v_s$, where $v_{j_0 s}$ is the frequency in slice $s$ of an arbitrarily chosen comb line with index $j_0$, and $\Delta v_s$ is the FSR of the interrogating comb in slice $s$. Furthermore, we assume that with the applied current ramp, $v_{j_0 s}$ and $\Delta v_s$ both change quadratically with $s$ (i.e., with time),

$$v_{js} = f(j, s) = (a_0 + a_1 s + a_2 s^2) + (j - j_0) \cdot (b_0 + b_1 s + b_2 s^2), \qquad (5)$$

where $a_0$ is the frequency of comb line $j_0$ in slice $s = 0$ (i.e., at $t = 0$), $a_1$ is the tuning slope for comb line $j_0$, and $b_0$ is the FSR of the interrogating comb in slice $s = 0$. For each absorption line, $l$, that we can identify in the rows of the absorbance matrix, we obtain three quantities: the centre frequency, $\hat{v}_l$, of the absorption line (e.g. from HITRAN); the comb line index, $j_l$, of the comb line that probed the absorption line $l$; and the slice index, $s_l$, at which the aforementioned comb line was at the center of the absorption line. With six or more absorption lines, the values of the polynomial coefficients in Eq. 5 can be determined through a nonlinear fit, i.e., by minimizing $\sum_l \left( (\hat{v}_l - f(j_l, s_l))^2 \right)$. In order to have good initial guesses for $a_0$ and $a_1$, for $j_0$ we choose a comb line that scans over at least two methane absorption lines. The starting value for $b_0$ is estimated by taking two methane lines probed by two different comb lines, $j_1, j_2$, and dividing the center frequency difference, $\Delta \hat{v}$, by the comb line index difference, $b_0 \approx \Delta \hat{v} / |j_1 - j_2|$. The initial values of the remaining fit parameters are set to zero.

The frequency axis in **Fig. 3** was established in this way through the known positions of 22 strong methane transitions. **Fig. 5** depicts the error in the transition center frequency of 55 methane transitions (including the 22 used for the calibration), defined as the difference between measured frequency (i.e., as determined from **Fig. 3**) and the position given in the HITRAN database [34]. The error for individual lines stays within 0.0015 cm$^{-1}$ and the rms value is $7 \times 10^{-4}$ cm$^{-1}$. The latter value corresponds to the spectral resolution of a classical Fourier transform spectrometer with an optical path difference of more than 10 m.

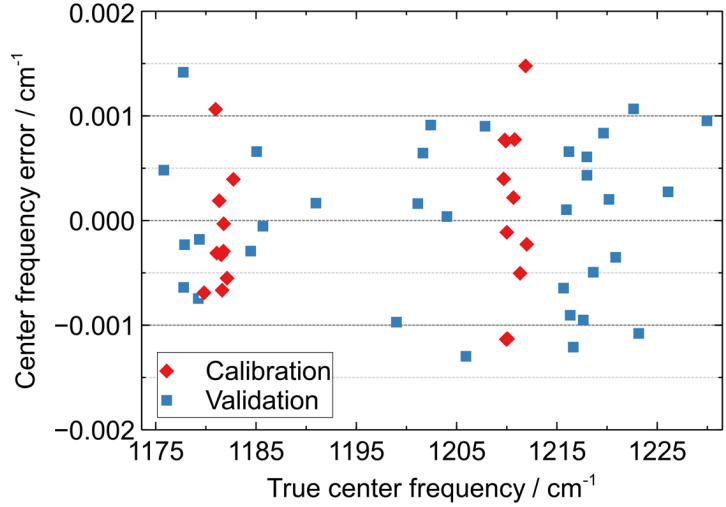

**Fig. 5.** Scatter plot of the center frequencies of 55 methane transitions. The transitions used for the frequency scale calibration are highlighted by red diamonds, the transitions used for validation by blue squares. Standard deviation of the center frequency errors: $7 \times 10^{-4}$ cm$^{-1}$.

*Noise characterization*

Here, we explore how co-averaging of multiple sweeps without any spectral smoothing (i.e., without any degradation of spectral resolution) affects the absorbance and phase noise. For this, we measured the amplitudes of the three beat notes (S, M, W) highlighted in **Fig. 2** in the first slice ($s = 0$) of 20,000 consecutive sweeps. In **Fig. 6**, the real and imaginary parts of the amplitude ratio (Eq. 1) for the three beat notes are shown. The phases of the three beat notes are not expected to be the same, except by coincidence. The magnitude of the ratio was normalized to unity. It is immediately obvious that the magnitude exhibits much better long-term stability than the phase. The phase drift over the 5 h duration of the measurement is approximately $\pi/2$ rad for all three beat notes.

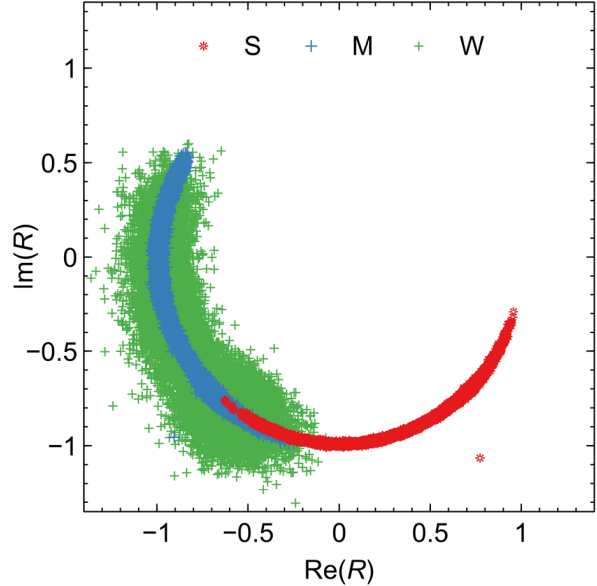

**Fig. 6.** Real and imaginary parts of the beat note amplitude ratio for the three beat notes (S, M, W) indicated in **Fig. 2**.

The logarithm of the magnitude and the phase of the amplitude ratio are shown in **Fig. 7**(a) and (b), respectively, as a function of sweep number. Fig. **7**(c) and (d) are Allan-Werle [35] plots for the two quantities above multiplied by $2\sqrt{2}$ and $\sqrt{2}$ to obtain the absorbance and phase rms, respectively (for more details see Appendix). The horizontal axis gives the measurement bandwidth, $\Delta f$, of the beat note amplitudes and is equal to 61 kHz/$n_s$, where 61 kHz is the resolution of the FFT used to measure the beat note amplitudes and $n_s$ is the number of sweeps. The behaviour of the absorbance (**Fig. 7**(c)) is that of white noise for up to 200 (S), 400 (M), and 1000 (W) co-averaged sweeps, corresponding to 3 min (S), 6 min (M), and 15 min (W) measurement duration, respectively. In the white noise regime, the absorbance rms is given by NEA $\cdot \sqrt{\Delta f}$, where NEA is the bandwidth-normalized noise equivalent absorbance. From **Fig. 7**(c) we observe that NEA $= 7 \times 10^{-5}$ Hz$^{-1/2}$ (S), $1.5 \times 10^{-4}$ Hz$^{-1/2}$ (M), and $1.0 \times 10^{-3}$ Hz$^{-1/2}$ (W). To date, to the best of our knowledge there are only a few publications using similar sources (quantum and interband cascade lasers) for dual comb spectroscopy. In two recent studies, NEA of $10^{-3}$ Hz$^{-1/2}$ [36] and $1.5 \times 10^{-3}$ Hz$^{-1/2}$ [14] were reported. Our data compare favorably to both examples, but neither of those studies employed interleaving techniques to reduce the spectral point spacing. Alternatively, we can derive the NEA from the residuals in **Fig. 4**. Since each data point is an average over 100 slices with a duration of 16.4 µs each, the bandwidth-normalized NEA becomes $7.4 \times 10^{-3} \cdot \sqrt{1.64 \text{ ms}} = 3.0 \times 10^{-4}$ Hz$^{-1/2}$. The amplitude of the beat note on which the absorption line in **Fig. 4** was measured is approximately 30 dB (relative to the detector noise floor, see **Fig. 2**), less than beat note M (~40 dB). In view of the difference in beat note amplitudes, the NEA determined from the residuals in **Fig. 4** is in coarse agreement with the derivation from the Allan-Werle plots for beat note M.

Next, we want to estimate the contribution of detector noise to the measured beat note amplitude ratios. Let $S_s, N_s$ be the amplitudes of any one of the three beat notes S, M, or W measured on the sample and normalizing detector, respectively, in slice $s$, and let $S_s^{\text{det}}, N_s^{\text{det}}$ be the Fourier coefficients in slice $s$ at the arbitrarily chosen frequency of 450 MHz when the laser beams are blocked. For each beat note S, M, or W, we then define $R_s = (\bar{S} + S_s^{\text{det}})/(\bar{N} + N_s^{\text{det}})$, with $\bar{S} = \langle |S_s| \rangle$ the average magnitude of the beat note (and similarly for $\bar{N}$). The ratio

$R_s$ is a simulation of the measured ratio if only detector noise were present. The derived absorbance rms plots are shown as dotted lines: from **Fig. 7**(c) it follows that the absorbance rms corresponding to beat notes with magnitude less or equal to that of M (which accounts for most of the beat notes) are detector noise limited. This is not the case for the strongest beat note S, where some excess noise of unknown origin is present.

The situation for the phase of the beat note amplitude ratio is very different. Once again, the presence of drift is immediately apparent in the phase plot (**Fig. 7**(b)). The drift is strongly correlated across all beat notes, and not solely the three beat notes examined here, and this might be caused by thermal drift, whereby the lengths of the various beam paths (**Fig. 1**) slowly change over the duration of the measurement (a 1 µm change in one of the beam paths would cause a phase shift of 0.8 rad). For the strongest beat note (S), the phase seems to be alternating between two states. The cause for that is as yet unclear. From **Fig. 7**(d), we find that there is virtually no benefit from co-averaging, except for the weakest beat notes, and co-averaging more than 100 sweeps is actually detrimental. If both absorbance and phase are of interest, the upper limit is thus 100 co-averaged sweeps, corresponding to 1.5 min measurement time. The phase rms for this duration is 20 mrad (independent of the beat note). Further work is required to pinpoint the cause of the phase drift and to mitigate it. Alternatively or additionally, phase correction algorithms [37,38] could also be explored.

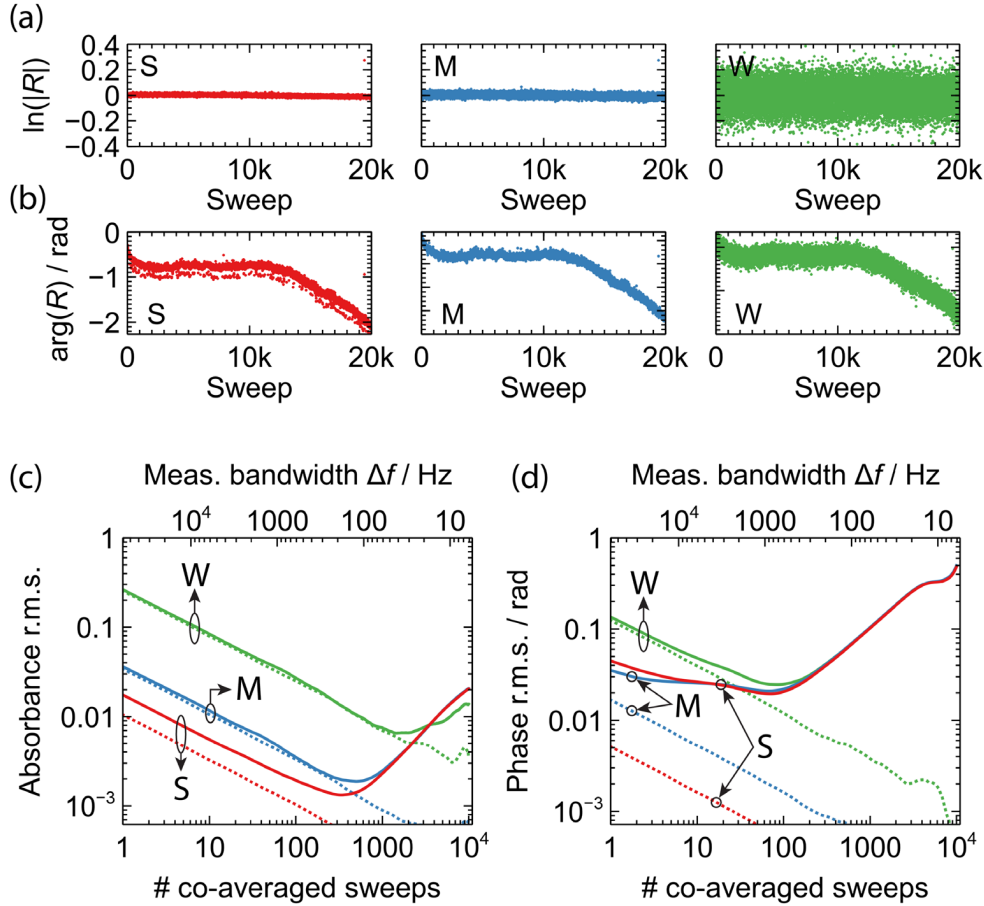

**Fig. 7.** Logarithm of magnitude (a) and phase (b) of beat note amplitude ratio for the three beat notes (S, M, W) highlighted in **Fig. 2**. Absorbance (c) and phase (d) rms for the same three beat notes as a function of number of co-averaged sweeps. Solid lines: measurements, dotted lines: simulations taking only detector noise into account. Measurement bandwidth: actual measurement bandwidth of the beat note amplitude ratios.

The noise equivalent power (NEP) of the photodetectors is specified by the manufacturer as 30 pW/√Hz averaged over the preamplifier bandwidth. With at most 250 µW of average optical power reaching the detectors, the shot noise amplitude spectral density is $\sqrt{2h\nu\bar{P}} = 3$ pW/√Hz ($h$: Planck's constant, $\nu = 3.6 \times 10^{13}$ Hz, $\bar{P} = 250$ µW), a factor of 10 below the detector NEP.

In summary, the absorbance rms is dominated by detector noise for all but the strongest beat notes, and there is a drift, which is likely to be of a thermal nature, that is particularly noticeable on the phase.

## 4. Summary

We have demonstrated an effective approach for gapless spectral coverage of the entire comb range of QCL-FCs. In a DCS configuration, synchronized laser current modulation has been used to sweep the spectra of both the interrogating and the local oscillator combs, maintaining the beat notes within the available detection bandwidth, and enabling a reduction of the spectral point spacing from 9.8 GHz (the FSR of the lasers) to below 1 MHz. The measurements covered 55 cm$^{-1}$ (1.65 THz) around 1200 cm$^{-1}$ (36 THz) without any gaps and with a resolution of 0.001 cm$^{-1}$ (30 MHz). We have also shown that the absorbance noise is given by the detector noise for most comb lines, and that co-averaging helps to reduce the noise.

### Appendix

*Derivation of absorbance and phase noise from beat note amplitude ratio noise*

A straightforward way of characterizing the absorbance noise of the spectrometer is to measure a beat note amplitude ratio twice under identical conditions, $R_1, R_2$, and then consider one to be the *background* measurement and the other the *sample* measurement. Each of $R_1$ and $R_2$ is assumed to be an average over $n$ individual measurements. The rms of the absorbance for a given value of $n$,

$$\sqrt{\langle A^2 \rangle} = 2\sqrt{\langle (\ln|R_2| - \ln|R_1|)^2 \rangle}, \qquad (5)$$

is a measure of the sensitivity of the spectrometer, where $\langle . \rangle$ stands for the expectation value over multiple repetitions of the measurements, $R_1, R_2$. With the definition, $\tilde{R}_i = \ln|R_i|$, $i = 1,2$, and recalling that the Allan variance of a quantity $\tilde{R}$ is $\sigma_{\tilde{R}}^2 = \frac{1}{2}\langle (\tilde{R}_2 - \tilde{R}_1)^2 \rangle$, we find

$$\sqrt{\langle A^2 \rangle} = 2\sqrt{2}\sigma_{\tilde{R}}. \qquad (6)$$

Hence, the plot of the absorbance rms as a function of number of averaged measurements, $n$, is simply the Allan-Werle plot for $\tilde{R} = \ln|R|$ multiplied by $2\sqrt{2}$.

For the phase we have,

$$\sqrt{\langle \psi^2 \rangle} = \sqrt{\langle (\arg R_2 - \arg R_1)^2 \rangle}. \qquad (7)$$

With the definition, $\tilde{R}_i = \arg R_i$, $i = 1,2$, we find

$$\sqrt{\langle \psi^2 \rangle} = \sqrt{2}\sigma_{\tilde{R}}. \qquad (8)$$

Hence, the plot of the phase rms as a function of averaged measurements, $n$, is simply the Allan-Werle plot for $\tilde{R} = \arg R$ multiplied by $\sqrt{2}$.

**FUNDING**

Swiss National Science Foundation Bridge Discovery program (40B2-0_176584).

**DISCLOSURES**

The authors declare no conflicts of interest.

**REFERENCES**

1. T. W. Hänsch, "Nobel lecture: Passion for precision," Rev. Mod. Phys. **78**, 1297–1309 (2006).
2. N. Picqué and T. W. Hänsch, "Frequency comb spectroscopy," Nat. Photonics **13**, 146 (2019).
3. A. Fried and D. Richter, "Infrared Absorption Spectroscopy," in *Analytical Techniques for Atmospheric Measurement*, D. E. Heard, ed., Markus W. Sigrist (Blackwell Publishing, 2006).
4. M. W. Sigrist, R. Bartlome, D. Marinov, J. M. Rey, D. E. Vogler, and H. Waechter, "Trace Gas Monitoring with Infrared Laser-Based Detection Schemes," Appl. Phys. B **90**, 289–300 (2008).
5. R. F. Machado, D. Laskowski, O. Deffenderfer, T. Burch, S. Zheng, P. J. Mazzone, T. Mekhail, C. Jennings, J. K. Stoller, J. Pyle, J. Duncan, R. A. Dweik, and S. C. Erzurum, "Detection of lung cancer by sensor array analyses of exhaled breath.," Am. J. Respir. Crit. Care Med. **171**, 1286–1291 (2005).
6. M. Phillips, "Breath Tests in Medicine," Sci. Am. **July**, 74–79 (1992).
7. F. Adler, K. C. Cossel, M. J. Thorpe, I. Hartl, M. E. Fermann, and J. Ye, "Phase-stabilized, 1.5 W frequency comb at 2.8–4.8 μm," Opt. Lett. **34**, 1330–1332 (2009).
8. S. A. Meek, A. Poisson, G. Guelachvili, T. W. Hänsch, and N. Picqué, "Fourier transform spectroscopy around 3 μm with a broad difference frequency comb," Appl. Phys. B **114**, 573–578 (2014).
9. F. C. Cruz, D. L. Maser, T. Johnson, G. Ycas, A. Klose, F. R. Giorgetta, I. Coddington, and S. A. Diddams, "Mid-infrared optical frequency combs based on difference frequency generation for molecular spectroscopy," Opt. Express **23**, 26814 (2015).
10. A. Khodabakhsh, V. Ramaiah-Badarla, L. Rutkowski, A. C. Johansson, K. F. Lee, J. Jiang, C. Mohr, M. E. Fermann, and A. Foltynowicz, "Fourier transform and Vernier spectroscopy using an optical frequency comb at 3–5.4 mum," Opt. Lett. **41**, 2541–2544 (2016).
11. P.-L. Luo, E.-C. Horng, and Y.-C. Guan, "Fast molecular fingerprinting with a coherent, rapidly tunable dual-comb spectrometer near 3 μm," Phys. Chem. Chem. Phys. **21**, 18400–18405 (2019).
12. C. Y. Wang, T. Herr, P. Del'Haye, A. Schliesser, J. Hofer, R. Holzwarth, T. W. Hänsch, N. Picqué, and T. J. Kippenberg, "Mid-infrared optical frequency combs at 2.5 μm based on crystalline microresonators," Nat. Commun. **4**, 1345 (2013).
13. M. Yu, Y. Okawachi, A. G. Griffith, N. Picqué, M. Lipson, and A. L. Gaeta, "Silicon-chip-based mid-infrared dual-comb spectroscopy," Nat. Commun. **9**, 1–6 (2018).
14. L. A. Sterczewski, J. Westberg, M. Bagheri, C. Frez, I. Vurgaftman, C. L. Canedy, W. W. Bewley, C. D. Merritt, C. S. Kim, M. Kim, J. R. Meyer, and G. Wysocki, "Mid-infrared dual-comb spectroscopy with interband cascade lasers," Opt. Lett. **44**, 2113–2116 (2019).
15. A. Hugi, G. Villares, S. Blaser, H. C. Liu, and J. Faist, "Mid-infrared frequency comb based on a quantum cascade laser," Nature **492**, 229–233 (2012).
16. G. Scalari, J. Faist, and N. Picqué, "On-chip mid-infrared and THz frequency combs for spectroscopy," Appl. Phys. Lett. **114**, 150401 (2019).
17. M. Yan, P.-L. Luo, K. Iwakuni, G. Millot, T. W. Hänsch, and N. Picqué, "Mid-infrared dual-comb spectroscopy with electro-optic modulators," Light Sci. Appl. **6**, e17076–e17076 (2017).
18. H. Timmers, A. Kowligy, A. Lind, F. C. Cruz, N. Nader, M. Silfies, G. Ycas, T. K. Allison, P. G. Schunemann, S. B. Papp, and S. A. Diddams, "Molecular fingerprinting with bright, broadband infrared frequency combs," Optica **5**, 727–732 (2018).
19. G. Ycas, F. R. Giorgetta, E. Baumann, I. Coddington, D. Herman, S. A. Diddams, and N. R. Newbury, "High-coherence mid-infrared dual-comb spectroscopy spanning 2.6 to 5.2 μm," Nat. Photonics **12**, 202–208 (2018).
20. Z. Chen, T. W. Hänsch, and N. Picqué, "Mid-infrared feed-forward dual-comb spectroscopy," Proc. Natl. Acad. Sci. **116**, 3454–3459 (2019).
21. F. Keilmann, C. Gohle, and R. Holzwarth, "Time-domain mid-infrared frequency-comb spectrometer," Opt. Lett. **29**, 1542–1544 (2004).
22. M. J. Thorpe, K. D. Moll, R. J. Jones, B. Safdi, and J. Ye, "Broadband Cavity Ringdown Spectroscopy for Sensitive and Rapid Molecular Detection," Science **311**, 1595–1599 (2006).
23. S. A. Diddams, L. Hollberg, and V. Mbele, "Molecular fingerprinting with the resolved modes of a femtosecond laser frequency comb," Nature **445**, 627–630 (2007).
24. C. Gohle, B. Stein, A. Schliesser, T. Udem, and T. W. Hänsch, "Frequency Comb Vernier Spectroscopy for Broadband, High-Resolution, High-Sensitivity Absorption and Dispersion Spectra," Phys. Rev. Lett. **99**, 263902 (2007).
25. I. Coddington, W. C. Swann, and N. R. Newbury, "Coherent Multiheterodyne Spectroscopy Using Stabilized Optical Frequency Combs," Phys. Rev. Lett. **100**, 013902 (2008).
26. J. Mandon, G. Guelachvili, and N. Picqué, "Fourier transform spectroscopy with a laser frequency comb," Nat. Photonics **3**, 99–102 (2009).
27. J. L. Klocke, M. Mangold, P. Allmendinger, A. Hugi, M. Geiser, P. Jouy, J. Faist, and T. Kottke, "Single-Shot Sub-microsecond Mid-infrared Spectroscopy on Protein Reactions with Quantum Cascade Laser Frequency Combs," Anal. Chem. **90**, 10494–10500 (2018).

28. I. Coddington, N. Newbury, and W. Swann, "Dual-comb spectroscopy," Optica **3**, 414–426 (2016).
29. S. Bartalini, S. Borri, P. Cancio, A. Castrillo, I. Galli, G. Giusfredi, D. Mazzotti, L. Gianfrani, and P. De Natale, "Observing the intrinsic linewidth of a quantum-cascade laser: Beyond the Schawlow-Townes limit," Phys. Rev. Lett. **104**, 1–4 (2010).
30. E. Baumann, F. R. Giorgetta, W. C. Swann, A. M. Zolot, I. Coddington, and N. R. Newbury, "Spectroscopy of the methane v3 band with an accurate midinfrared coherent dual-comb spectrometer," Phys. Rev. A **84**, 062513 (2011).
31. B. Spaun, P. B. Changala, D. Patterson, B. J. Bjork, O. H. Heckl, J. M. Doyle, and J. Ye, "Continuous probing of cold complex molecules with infrared frequency comb spectroscopy," Nature **533**, 517–520 (2016).
32. P. Jouy, J. M. Wolf, Y. Bidaux, P. Allmendinger, M. Mangold, M. Beck, and J. Faist, "Dual comb operation of λ ∼ 8.2 µm quantum cascade laser frequency comb with 1 W optical power," Appl. Phys. Lett. **111**, 141102 (2017).
33. F. J. Harris, "On the use of windows for harmonic analysis with the discrete Fourier transform," Proc. IEEE **66**, 51–83 (1978).
34. I. E. Gordon, L. S. Rothman, C. Hill, R. V. Kochanov, Y. Tan, P. F. Bernath, M. Birk, V. Boudon, A. Campargue, K. V. Chance, B. J. Drouin, J.-M. Flaud, R. R. Gamache, J. T. Hodges, D. Jacquemart, V. I. Perevalov, A. Perrin, K. P. Shine, M.-A. H. Smith, J. Tennyson, G. C. Toon, H. Tran, V. G. Tyuterev, A. Barbe, A. G. Császár, V. M. Devi, T. Furtenbacher, J. J. Harrison, J.-M. Hartmann, A. Jolly, T. J. Johnson, T. Karman, I. Kleiner, A. A. Kyuberis, J. Loos, O. M. Lyulin, S. T. Massie, S. N. Mikhailenko, N. Moazzen-Ahmadi, H. S. P. Müller, O. V. Naumenko, A. V. Nikitin, O. L. Polyansky, M. Rey, M. Rotger, S. W. Sharpe, K. Sung, E. Starikova, S. A. Tashkun, J. V. Auwera, G. Wagner, J. Wilzewski, P. Wcisło, S. Yu, and E. J. Zak, "The HITRAN2016 molecular spectroscopic database," J. Quant. Spectrosc. Radiat. Transf. **203**, 3–69 (2017).
35. P. Werle, R. Mücke, and F. Slemr, "The Limits of Signal Averaging in Atmospheric Trace-Gas Monitoring by Tunable Diode-Laser Absorption-Spectroscopy (TDLAS)," Appl. Phys. B **57**, 131–139 (1993).
36. J. Westberg, L. A. Sterczewski, F. Kapsalidis, Y. Bidaux, J. M. Wolf, M. Beck, J. Faist, and G. Wysocki, "Dual-comb spectroscopy using plasmon-enhanced-waveguide dispersion-compensated quantum cascade lasers," Opt. Lett. **43**, 4522–4525 (2018).
37. N. B. Hébert, J. Genest, J.-D. Deschênes, H. Bergeron, G. Y. Chen, C. Khurmi, and D. G. Lancaster, "Self-corrected chip-based dual-comb spectrometer," Opt. Express **25**, 8168 (2017).
38. L. A. Sterczewski, J. Westberg, and G. Wysocki, "Computational coherent averaging for free-running dual-comb spectroscopy," Opt. Express **27**, 23875–23893 (2019).